\documentclass[journal]{IEEEtran}
\usepackage{cite}
\usepackage{graphicx}
\usepackage{array}
\usepackage{tikz}
\usepackage{tikz-3dplot}
\usetikzlibrary{shapes.geometric}
\usetikzlibrary{decorations.pathreplacing}
\usetikzlibrary{positioning}
\usetikzlibrary{patterns}
\usepackage{pgfplots}
\usepackage{pstool}  
\usepackage{mathtools}
\usepackage[linesnumbered,algoruled,boxed,lined]{algorithm2e}
\usepackage[absolute]{textpos}
\usepackage{threeparttable}
\usepackage[strict]{changepage}
\usepackage{amssymb}
\usepackage{comment}
\usepackage{bbm}
\usepackage{hhline}
\usepackage{lipsum} 
\usepackage{enumitem}
\usepackage{soul}
\usepackage{xcolor}

\usepackage{amsthm}

\ifCLASSINFOpdf
\else
\fi
\usepackage{amsmath}
\ifCLASSOPTIONcompsoc
  \usepackage[caption=false,font=normalsize,labelfont=sf,textfont=sf]{subfig}
\else
  \usepackage[caption=false,font=footnotesize]{subfig}
\fi
\newcommand{\argmax}{\operatorname{arg\, max\,}}

\pgfplotsset{compat=1.14} 

\begin{document}
%
\title{Unsupervised Learning in Next-Generation Networks: Real-Time Performance Self-Diagnosis}
%
\author{Faris B.~Mismar,~\IEEEmembership{Senior Member,~IEEE},
       and~Jakob Hoydis,~\IEEEmembership{Senior Member,~IEEE}
\thanks{The authors are with Nokia USA and Nokia Bell Labs. Email: faris.mismar@nokia.com and jakob.hoydis@nokia-bell-labs.com.}
}%

\maketitle

\begin{abstract}
This letter demonstrates the use of unsupervised machine learning to enable performance self-diagnosis of next-generation cellular networks.  We propose two {simplified} applications of unsupervised learning that can enable real-time performance self-diagnosis on edge nodes such as the radio access network intelligent controller (RIC).  The first application detects anomalous performance and finds its root cause of faults, configuration, or network procedure failures.  The second application uses clustering to learn the relationship between two performance measures.  Our proposed applications run in near-constant time complexity, making them, combined with subject-matter expertise validation, suitable real-time RIC applications for network diagnosis.

\end{abstract}


%
\IEEEpeerreviewmaketitle

\section{Introduction} 
%
%
%
%



\begin{figure}[!t]
\centering
\resizebox{0.48\textwidth}{!}{\begin{tikzpicture}[style=thick, node distance=3cm, scale=2, >=latex]
    \node [coordinate, name=input] {};
	\node [rectangle, right of=input,
		text width=6em, text centered, minimum height=2em] (XM) {};
	\node [rectangle, draw, right of=XM,
		text width=6em, text centered, minimum height=2em] (CM) {Configuration};
	\node [rectangle, draw, right of=CM,
		text width=6em, text centered, minimum height=2em] (FM) {Fault};
    \node [rectangle, draw, right of=FM,
		text width=6em, text centered, minimum height=2em] (PM) {Performance};
    \node [rectangle, draw, below of=XM, node distance=2.5em,
		text width=6em, text centered, minimum height=2em] (IC) {Edge Node};
    \node [rectangle, draw, right of=IC, node distance=6cm,
		text width=18em, text centered, minimum height=2em] (BS) {Next-Generation Base Station};
    
     \node [rectangle, fill=gray!20, draw, below of=IC, node distance=3em,
 		text width=6em, text centered, minimum height=2em] (unsup) {Unsupervised \\ Learning};
		
	\node [rectangle, fill=gray!40, draw, below of=BS, node distance=3em,
 		text width=6em, text centered, minimum height=2em] (rca) {Root Cause \\ Analysis};
	
	\node [rectangle, fill=gray!40, draw, right=of rca, node distance=3em,
 		text width=6em, text centered, minimum height=2em] (rd) {Relationship Discovery};

     \draw [draw,<->] (IC) -- node {} (BS);
		

	
    
    

\end{tikzpicture}%
}%
\caption{Next-generation base station connected to the edge, such as the RAN intelligent controller (RIC), which can learn from the various data sources.}
\label{fig:overall}
\end{figure}
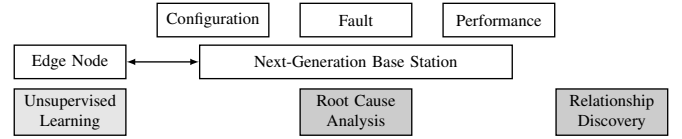

 
The proliferation of machine learning (ML) in wireless communications applications in the fifth generation of wireless communications (5G) and beyond motivates the introduction of general purpose computational hardware to the transmit and receiving communication ends \cite{mismar}.  In a cellular system, these ends are the user equipment (UE) and the base station (BS).  Industry standards have enabled edge storage and computation in next-generation radio access networks (RAN), such as the Open-RAN RAN intelligent controller (RIC), where network data operations can be performed either real-time or non-real-time \cite{o-ran}.  Next-generation RAN data can be broadly classified into fault management (FM), performance management (PM), and configuration management (CM) data.  While PM data is often aggregated in time intervals, FM and CM data are instantaneous since they are event-driven. PM data can be further classified as counters and key performance indicators (KPIs), which are derived from these counters as \textit{important} measures of performance of the services offered by the network {and are defined in the standards} \cite{3gpp28554}.  {An architectural sketch of the BS, edge node, and network data is shown in Fig.~\ref{fig:overall}.}

While the level of detail of PM counters is left to vendors, PM counters contain RAN-reported causes for RAN procedure failures (e.g., call setup, call release, handovers) as well as measurements related to the uplink and downlink across protocol stack layers.  Combined together with the changes and faults that occur on a BS, an \textit{observability space} or \textit{universe} is formed {per BS}, {which includes confounding variables} and therefore causal analysis can be performed.  The larger this observability space is (i.e., in terms of the number of elements captured), the more causes of performance degradation can be considered and the more pairs can be constructed for relationship discovery.  We use the term \textit{feature} to refer to any of these elements.

The use of unsupervised learning involves inferring patterns in data without a known true reference.  Hence, one known challenge in employing unsupervised learning is that it requires a subject-matter expert intervention.  

{Another challenge is practicality, where \textit{simplification} of the problem statement helps reduce required implementation efforts and thus the monetary expenses involved in execution.}  In this letter, we show how to overcome these challenges and enable two different use cases related to automation of performance self-diagnosis in next-generation RAN.

Prior work related to automating RAN performance monitoring and self-optimizing can be found in \cite{stoica,8758212,8030498,8939988}. Automating diagnosis of RAN was studied in \cite{stoica}, where individual models of \textit{specific} KPIs were devised using supervised learning.  However, as next-generation networks continue to evolve, the number and complexity of KPIs will also evolve, posing a challenge to such models. A random forest classifier was used to construct an influence matrix, based on feature importance, which may indicate correlation but not causality. Learning probability densities of performance data to detect KPI degradation was studied in \cite{8939988}.  However, the focus was for PM data only (i.e., no CM or FM data) and no attribution to a root cause---two aspects covered in our letter.

This letter makes two specific contributions:
\begin{enumerate}
\item Create a simplified {approach} to automate RAN self-diagnosis in real-time using a single class of learners and show two applications of this {class of learners}.
\item Create a transformation that enables combining different data sources with different natures (i.e., event driven vs.\ aggregated) to a single design matrix.
\end{enumerate}

To this extent, we demonstrate two  applications: 1) root cause analysis using \textit{anomaly detection} and exclusion of features known to the subject-matter expert not to cause the anomaly despite their strong correlation and 2) relationship discovery using \textit{clustering} and aggregation to derive a tabular lookup structure as a function of two RAN data features.

\section{Network Model}

In this letter, we consider a system composed of multiple BSs.  Each BS has a group of UEs in its respective service area.  We impose no constraints on the services, direction, frequency bands, antenna count, or mobility on either the UEs or the BSs.  This is justified since an observability space contains information about all these areas.

\textbf{Protocol stack:} with a focus on the RAN air interface (i.e., between the BS and the UE), the data generated by the BS can either 1) have an impact on the network at different layers in the protocol stack (in the case of faults or configuration) or 2) observe the performance of these different layers (in the case of performance).

\textbf{Edge nodes:} next-generation networks introduce the concept of the ``edge node.''  This edge node resides alongside the BS and has the capability to collect the FM, CM, and PM data generated by the BS. It further can run code that implements the contributions presented in this letter. This edge node can be the RIC as mentioned earlier, which enables near real-time and extensible capability through modular software plug-ins, such as the two applications presented in this letter.


\vspace*{-0.5em}
\section{Primer on Unsupervised Learning}\label{sec:unsupervised}
This section explains the relevant details of the two unsupervised learning algorithms used in this letter.%
\vspace*{-1em}
{
\subsection{Density Based Spatial Clustering}\label{sec:dbscan}%
``Density based spatial clustering of applications with noise'' (DBSCAN) \cite{dbscan} is an unsupervised learning algorithm that 
groups points with high neighbors count as high ``density.'' It marks single points in low-density regions as \textit{anomalous}.

\textbf{Training:} DBSCAN identifies a set of points in the dataset with adequate neighbors and constructs clusters from them based on their Euclidean distance from these neighbors.  A cluster is formed if it has a minimum number of points.  Then, the algorithm expands the clusters formation by computing the distance and cluster size for each neighboring point for all points in the dataset.  An anomalous point has a low number of neighbors at a minimum distance from itself.

\textbf{Hyperparameters:} $\epsilon > 0$ specifies how close points should be to each other to be considered a part of a cluster (distance below $\epsilon$ means that these points become neighbors). The minimum number of samples (minPts) define the least number of points in a cluster that can form a ``dense'' region.  These hyperparameters can be tuned using grid search.

\textbf{Outcome:} DBSCAN stops once it has visited all points in the dataset.  A binary column vector $\mathbf{x_{1b}}$ is produced indicating whether a value from $\mathbf{x}_1$ is anomalous (1) or not (0).
}
\vspace*{-1em}
\subsection{One-Dimensional $k$-Means Clustering}
The $k$-means clustering algorithm \cite{1056489} partitions the data into distinct clusters, minimizing the sum of the square of the Euclidean distance of these data points with respect to defined cluster centroids.  A label is assigned to each data point based on its closest centroid.

\textbf{Initialization:} $k$-means has a vanilla implementation that initializes the cluster centroids at random \cite{scikit-learn}.  However, we use a modification that allows us to deterministically initialize and fix the coordinate of these centroids on a line---hence one-dimensional.  These individual coordinates need to span the entire range of the training data vector {to avoid losing information}.  We {optionally} choose these coordinates to be equi-distant {for better smoothing performance and to avoid losing information in missed data points between coordinates.}

\textbf{Training:} the training data $\mathbf{x}_1$ is a column vector corresponding to one of the data features with $n$ elements. The training phase iterates to find the closest centroid to each element using Euclidean distance.  It strictly converges when the the labels in two adjacent iterations do not change.  This runs in $\mathcal{O}(nk)$ \cite{scikit-learn}.

\textbf{Hyperparameters:} since we choose equi-distant centroids that span the range of the training data, the number of clusters $k$ is tuned based on how wide this range is: a small $k$ with a wide range {causes information loss due to assigning dispersed data to one centroid}.  Also, a large $k$ may cause some of the clusters to be associated with no data points (i.e., pigeonholed).  We fix the number of iterations in the training phase.  

\textbf{Outcome:} once the $k$-means algorithm has stopped, each training data point $[\mathbf{x_1}]_i$ then gets a centroid label $[\mathbf{{s}}]_i\in\{1,\ldots,k\}$.  The output thus has the form $[\mathbf{x}_1, \mathbf{{s}}]$, where $\mathbf{{s}}\in\mathbb{Z}_+^n$ is also a column vector.

\section{Data Construction}\label{sec:dataprep}
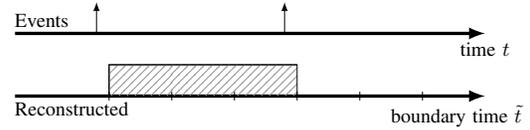
\begin{figure}[!t]
\centering
\resizebox{0.43\textwidth}{!}{\begin{tikzpicture}[node distance=4.5cm, auto, scale=1]
\tikzstyle{every node}=[font=\small]
\linespread{0.9}

\node[text width=2cm] at (0.5,0.2) {\small Events};
\draw[line width=1.5, ->, >=latex] (-0.5,0) -- (7,0) node[below] {time $t$};
\node[text width=2cm] at (0.5,-1.2) {\small Reconstructed};
\draw[line width=1.5, ->, >=latex] (-0.5,-1) -- (7,-1) node[below, text width=3cm] {boundary time $\tilde{t}$};

\foreach \x in {1,4}
    \draw[->, >=latex] (\x-0.2,0) -- ++(0,0.5); 

\foreach \x in {1,...,6} {
    \draw (\x,-1cm-2pt) -- (\x,-1cm+2pt) node [below] {};
}

\draw[line width=0.25] (1,-1) foreach \i in {1,4}{-|(\i,-0.5)};

\draw[pattern=north east lines, pattern color=gray!80] (1,-1) rectangle (4,-0.5);
\end{tikzpicture}}%
\vspace*{-1em}\caption{Reconstructing an event-driven data feature (e.g., configuration or fault) through binning and integration.}
\label{fig:reconstruction}
\end{figure}%

While obtaining data for the mere purposes of authoring this letter posed a significant challenge, combining readily available RAN data comprising PM, CM, and FM comes with another challenge.  On the one hand, the PM data reports $\mathbf{P} \coloneqq [\mathbf{p}_1, \mathbf{p}_2, \ldots \mathbf{p}_r], \mathbf{P}\in\mathbb{R}^{m\times r}$ are an \textit{aggregation} of counters and KPIs with \textit{periodicity} $\Delta T > 0$ and a time window $W > 0$ where $m \coloneqq \lceil W/\Delta T \rceil$.  On the other hand, the CM data $\mathbf{C}$ and FM data $\mathbf{F}$ are \textit{event-driven} (i.e., once a hardware fault appeared or disappeared or a parameter change was made).  Given that these events do not have to happen at a periodicity boundary, a step of time binning (i.e., rounding to a uniformly spaced boundary) then integration of these discrete events take place first, as shown in Fig.~\ref{fig:reconstruction}.  To explain this, {consider if an alarm $a$ occurred on a BS at time $t_1$, then we can represent this as $f^{(a)}(t) \coloneqq \delta[t - \tilde{t}_1]$, where $\delta[\cdot]$ is the unit impulse function.  Similarly, when the alarm $a$ is cleared at time $t_1 < t_2 \le +\infty$, then we have $f^{(\lnot a)}(t) \coloneqq \delta[t - \tilde{t}_2]$.  Here we use the notation $\tilde{t}_i$ as a binned time $t_i$ with respect to the periodicity $\Delta T$ (i.e., $\tilde{t}_i \coloneqq [t_i / \Delta T] \Delta T$).  This enables us to construct a rectangular pulse $F^{(a)}(t)$ the duration of which is the difference in the binned time it took the alarm to be cleared (i.e., $\tilde{t}_2 - \tilde{t}_1$).  This pulse value is 1 or 0 whether an alarm occurred or was cleared within $t$.  It is {similar} to derive the CM signal $C^{(b)}(t)$ following the same logic {used for FM} where parameter $b$ changes from $\alpha$ to $\beta$ at times $t_1$ and $t_2$.

For a group of faults $\mathcal{F} \coloneqq \{1,\ldots,p\}$, we can construct several column vectors $\mathbf{f}^\prime_x \coloneqq [F^{(x)}(t)]_{t = 0}^{m-1}, x\in\mathcal{F}$.  Similarly, for a group of parameter changes $\mathcal{P} \coloneqq \{1,\ldots, q\}$, the parameter change column vectors can be written as $\mathbf{c}^\prime_y \coloneqq [C^{(y)}(t)]_{t = 0}^{m-1}, y\in\mathcal{P}$.}

The edge node of a given BS can now construct a design matrix $\mathbf{M}\in\mathbb{R}^{m\times n}$, $\mathbf{M} \coloneqq [\mathbf{p}_1, \ldots, \mathbf{p}_r, \mathbf{f}^\prime_1\ldots,\mathbf{f}^\prime_p, \mathbf{c}^\prime_1, \ldots, \mathbf{c}^\prime_q]$. {$\mathbf{M}$ contains in each row PM, FM, and CM values for a particular time period.} For this matrix, the number of rows $m \coloneqq \lceil W/\Delta T\rceil$ as shown earlier. The number of columns $n$ is the sum $n \coloneqq r + p + q$.





\section{Root Cause Analysis}

An RCA algorithm in RAN has to answer two questions: 1) Which configuration change, fault, or RAN procedure failure caused a degradation in a given KPI?  2) How certain can we be {about the answer compared to that of a subject-matter expert?}

\textbf{Unsupervised learning:} to answer the first question, we construct a two-dimensional dataset comprising the degraded KPI and another column from the design matrix $\mathbf{M}$ and apply the {DBSCAN} algorithm on this dataset as an anomaly detection function $f\colon \mathbb{R}\rightarrow \{0,1\}$.  This converts the design matrix $\mathbf{M}$ to a binary matrix $\mathbf{M_b} \coloneqq [\mathbf{m_b}_1, \ldots \mathbf{m_b}_n], \mathbf{m_b}_i\in\{0,1\}^m, i \in \{1,2,\ldots, n\}$, where each element is zero if not anomalous in its respective column and equal to one if anomalous. Let the ${v}$-th column in $\mathbf{M}$ be the column of the degraded KPI, where $\mathbf{m_b}_{v}$ is set whether the KPI is above or below an \textit{absolute} threshold.  {The hyperparameters of DBSCAN are optimized through grid search cross validation.}

\textbf{Correlation:} The next step in the approach is to study the correlation between $\mathbf{m_b}_{v}$ and all other columns.  We apply a modified Pearson's Phi correlation coefficient $\phi(\cdot,\cdot)$ pair-wise {(where it is set to zero for the $v$-th column since a KPI degradation cannot be its own root cause)} and rank the columns based on their correlation \textit{absolute} value {where it is defined}.  This circumvents the problem of high data (i.e., data where higher values are considered good such as throughput) and low data (i.e., data where lower values are considered good such as the drop rate).  The pair-wise correlation enables us to construct a column vector of correlations $\mathbf{r_{v}}\in\mathbb{R}_+^n$.



\textbf{Causality:} given that a correlation does not imply causality, a causality filter $\mathbf{{u}} \in\{0,1\}^n$ is applied to the output, the entries of which map to the same order of the counters and KPIs in $\mathbf{r_{v}}$. This reduces the number of elements based on whether a highly correlated column in the data is \textit{known} by a subject-matter expert to also be a \textit{cause}. The subject-matter expert builds this vector only once for every KPI. The element-wise product of the two vectors $\mathbf{g} \coloneqq \mathbf{{u}} \odot \mathbf{r_{v}}$ 
finally enables writing the RCA as:
\begin{equation}
i^\ast \coloneqq \underset{i\in[1,n]}{\argmax}\; [\mathbf{g}]_i.
\label{eq:rca}
\end{equation}

We use the value of this $i^*$-th element in $\mathbf{r_{v}}$ as a measure of certainty that the $i$-th column in $\mathbf{M}$ is the root cause of the degradation of the ${v}$-th column, which is the answer of the second question.  This could be a CM, FM, or PM data column, as explained in this section.  We show the major steps of our proposed RCA algorithm in Algorithm~\ref{alg:algorithm_rca}.

\textbf{Hyperparameter tuning:} the hyperparameters are {$\epsilon$ and minPts. These hyperparameters are tuned using grid search to maximize $\vert\phi(\cdot,\cdot)\vert$, as discussed in Section~\ref{sec:unsupervised} and Section~\ref{sec:simulation}.}

\textbf{Run-time complexity:} the run-time complexity of the proposed RCA {using DBSCAN} is in $\mathcal{O}(n)$, which we validate in Section~\ref{sec:simulation}. For a fixed $n$, it becomes $\mathcal{O}(1)$.

\section{Relationship Discovery}

Relationship discovery learns a tabular estimation that describes a \textit{monotonic} relationship of one PM data feature (i.e., the independent variable) to another (i.e., the target variable) in the presence of heteroscedasticity and often incomplete or missing data in either one of the two data features.

The idea of generating such relationships from data is not new.  In essence, it converts a parametric equation of two quantities into a standard form through simultaneously solving both quantities for the parameter variable. In RAN performance data, the parameter variable is time (e.g., during the time window $W$ as we showed in Section~\ref{sec:dataprep}).

\begin{algorithm}[!t]
    \caption{Root Cause Analysis}
    \label{alg:algorithm_rca}
    \DontPrintSemicolon
    \KwIn{RAN data design matrix $\mathbf{M}$, $v$ the column of degraded KPI, and the causality filter $\mathbf{{u}}$.}
    \KwOut{Column name of the root cause of degradation and a score.}
    Sequentially pick a column from $\mathbf{M}$ as $\mathbf{m}_i, i\neq v$\;
    Run DBSCAN on $[\mathbf{m}_i, \mathbf{m}_v]$ using grid search\;
    Store the binary output in $\mathbf{M_b}\in\{0,1\}^{m\times n}$\;
    Compute $[\mathbf{r_{v}}]_i \coloneqq \vert \phi(\mathbf{m_b}_i, \mathbf{m_b}_{v})\vert$ using Pearson's Phi\;
    $\mathbf{g} = \mathbf{{u}} \odot \mathbf{r_v}$\;
    $i^\ast \coloneqq \argmax_i\; [\mathbf{g}]_i$\;
    \Return \{$i^\ast, [\mathbf{r_v}]_{i^\ast}$\}\;
\end{algorithm}%

\begin{algorithm}[!t]
    \caption{Relationship Discovery}
    \label{alg:algorithm_rd}
    \DontPrintSemicolon
    \KwIn{Two performance features $\mathbf{p}_X$ and $\mathbf{p}_Y$, number of clusters $k$, a threshold $\gamma$ and a choice of aggregation.}
    \KwOut{Lookup tabular structure
    $\mathbf{D}\in\mathbb{R}^{k\times 2}$ and an optional smooth output}
    Initialize $k$-means centroids $\mathbf{k}$\;
    Using $k$-means, obtain $\mathbf{p}_X^{(k)} \coloneqq \mathsf{kmeans}(\mathbf{p}_X; \mathbf{k})$\;
    Construct $\mathbf{x} \coloneqq \mathsf{centroid}(\mathbf{p}_X^{(k)}; \mathbf{k})$\;
    Construct $\mathbf{y}$ from $\mathbf{p}_Y$ using \eqref{eq:y_from_py}\;
    $\mathbf{D} \coloneqq [\mathbf{x}, \mathbf{y}]$ \;
    Apply the smoothing function $\mathbf{y}\mapsto \mathbf{\tilde y}$\;
    \Return $\{\mathbf{D}, \mathbf{\tilde y}\}$\;
\end{algorithm}%

The reason we revisit this idea in PM data for RAN is three-fold: 1) simultaneously solving for two equations requires the existence of such equations, which is not the case in PM data in RAN, where only counters and KPIs are collected and tabulated, 2) known closed-form solutions, if exist, may be derived for ideal cases or scenarios that do not hold in a practical network, and 3) not all the values belonging to the domain of the function to be learned are always present in the PM data during the time window of interest $W$ (i.e., discontinuity).  We show how our proposed technique behaves under these three challenges.



\textbf{Closed-form solutions:} consider two PM data counters or KPIs such as the uplink spectral efficiency (SE) and the uplink signal-to-interference-plus-noise ratio (SINR) as measured on the uplink channel.  If we assume perfectly known channel state information (CSI) at the UE, then due to Shannon:
\begin{equation}
    \mathsf{SE} = \log_2(1 + \mathsf{SINR}),
    \label{eq:shannon}
\end{equation}
which is a closed-form and a theoretical upper limit for the SE.  For the downlink the channel quality indicator \cite{3gpp38214} can be used as an integer proxy for the downlink SINR.  However, in practical networks with interference, link adaptation, and imperfect knowledge of the CSI during a given time window $W$, where a relationship curve is desired, this known closed-form formula becomes less applicable. In fact, there are other RAN performance relationships that may not have a known closed-form such as the downlink traffic volume vs.\ the number of active users.

\textbf{Unsupervised learning:} we propose {the use of one-dimensional $k$-means clustering with the cluster centroids that are equi-distant and span the range of the independent variable.} We then aggregate the values of the dependent variable of the samples belonging to {each cluster of the $k$ clusters.}

To explain this, let us use $\mathbf{p}_X$ and $\mathbf{p}_Y$ as the independent and target variables respectively.  These can be any of the column vectors from the PM data $\mathbf{P}$ as motivated in Section~\ref{sec:dataprep}. Then for a {cluster size $k > 0$, we initialize the clusters centroids $\mathbf{k} \coloneqq [\min(\mathbf{p}_X),\ldots, \max(\mathbf{p}_X)],\mathbf{k}\in\mathbb{R}^k$.  Since $k$-means is effectively a function that accepts $\mathbf{p}_X$ and returns the label of the cluster that the data point belongs to, we write $\mathbf{p}_X^{(k)} \coloneqq \mathsf{kmeans}(\mathbf{p}_X; \mathbf{k})$}.

\textbf{Aggregation:} given the labels $\mathbf{p}_X^{(k)}$, we can aggregate all the data points from $\mathbf{p}_Y$ that belong to a given label $j$, effectively partitioning the $\mathbf{p}_X$-$\mathbf{p}_Y$ space:
\begin{equation}\label{eq:y_from_py}
[\mathbf{y}]_j \coloneqq \begin{cases}
    \mathsf{aggregate}(\mathbf{p}_Y\,\vert\,\mathbf{p}_X^{(k)}=j), & \#(\mathbf{p}_X^{(k)} = j) > \gamma \\
    \text{NaN} & \text{otherwise}
\end{cases},
\end{equation}%
where $j \in [0,k-1]$, $k \le m$, and $\gamma \gg 0$ is a threshold of significant sample size.  This aggregation step (e.g., average or max) in constructing $\mathbf{y}\in\mathbb{R}^k$ helps reduce the impact of heteroscedasticity on the data. For clusters with inadequate or no samples (i.e., NaN or missing data), an imputation technique in $\mathbf{y}$ (e.g., replacing with last value) can work.  {$k$-means converts the labels to their respective centroids $\mathbf{x} = \mathsf{centroid}(\mathbf{p}_X^{(k)}; \mathbf{k})$, which enables constructing a lookup tabular structure $\mathbf{D} \coloneqq [\mathbf{x}, \mathbf{y}]$ as} a data-driven function.  The final step is to pass $\mathbf{y}$ through a smoothing function $\mathbf{y}\mapsto \mathbf{\tilde y}$ the choice of which can be left to the subject-matter expert to enhance the function for visual presentation.  We show the major steps of our proposed technique in Algorithm~\ref{alg:algorithm_rd}.

\textbf{Hyperparameter tuning:} there are three hyperparameters here: 1) the number of clusters $k$, 2) $\mathsf{aggregate}(\cdot)$, and 3) the threshold $\gamma$. {We fix the random generator seed to $0$.}

While our proposed technique attributes the target variable as a function of an independent variable, it is not considered a supervised learning regression. Here is why: 1) regression uses the true value of the target variable to minimize a cost function (i.e., typically the difference between the true and predicted values of the target variable), a function which cannot be defined since we only have predicted values and 2) the variance of the difference between the true and predicted target variable must be constant across observations in regression (i.e., homoscedasticity), while we account for heteroscedasticity through aggregation, as we stated earlier.

\textbf{Run-time complexity:} the run-time of the relationship discovery for a fixed $k$ is $\mathcal{O}(n)$.  We validate this claim in Section~\ref{sec:simulation}.%

\vspace*{-1em}
\section{Simulation}\label{sec:simulation}%

\vspace*{-1em}
\begin{table}[!t]
\centering
\setlength\doublerulesep{0.5pt}
\caption{Simulation settings}
\vspace*{-1em}
\label{table:simulation}
\begin{tabular}{ lll } 
\hhline{===}
Algorithm & Parameter & Setting \\
 \hline
RCA & Normalized window size $W / \Delta T$ & \{5, 120\} \\
 & DBSCAN $\epsilon$ & \{0.1, 0.3, 0.5\} \\
\hline
Relationship  & $\mathsf{aggregate(\cdot)}$ & \{max, average\} \\
Discovery& Threshold $\gamma$ & \{100, 200\}\\
\hhline{===}
\end{tabular}
\end{table}%

\begin{table}[!t]
\centering
\setlength\doublerulesep{0.5pt}
\caption{RCA Performance}
\vspace*{-1em}
\label{table:performance}
\begin{tabular}{ cc | cc }
\hhline{====}
\multicolumn{2}{c|}{Scenario A} &  \multicolumn{2}{c}{Scenario B}\\
\hline
$W/\Delta T$ & Accuracy & $W/\Delta T$ & Accuracy \\
\hline
5 & {0.8913} & 120 & {0.9932} \\
\hhline{====}
\end{tabular}
\end{table}%

\subsection{Setup}
We show the simulation parameters in Table~\ref{table:simulation}. Using a mobile operator dataset comprising PM, FM, and CM data, we construct a design matrix $\mathbf{M}$ as in Section~\ref{sec:dataprep} and per BS with an hourly periodicity (i.e., $\Delta T = 1$ row per hour). This matrix has $n = 266$ columns of counters, KPIs, alarms, and parameter changes. {The KPI we study ($\mathbf{m_b}_v$) is the bearer drop rate.  An example of counter building this KPI ($\mathbf{m_b}_q, q\neq v$) is the number of releases due to connection loss with UE}.

For the proposed automated RCA method, we use a time window $W$ of 5 working days. Therefore, we have $W/\Delta T = 5 \times 24 / 1 = 120$ rows of data per BS.  We {set minPts to 5 and} search over $\epsilon$ as shown in Table~\ref{table:simulation} to maximize $\vert\phi(\cdot,\cdot)\vert$ and simulate two scenarios: Scenario A uses 
only the hour of busiest traffic per working day (the busy hour) and therefore has only $5$ rows per BS. Scenario B uses 
all hours per day.

For the proposed relationship discovery method, we also use hourly data to generate relationship plots but for one BS and a period of one month.  This is necessary to capture some seasonality and unusual traffic patterns in the data (and hence heteroscedasticity).  We set the number of clusters $k = 30$ and show scatter plots based on reported PM.  Similarly, we simulate two scenarios: Scenario A has a known closed-form solution and Scenario B is not known to have a closed-form solution. As a benchmark for both scenarios, we use a regression model with adequate complexity: a fully connected deep neural network (DNN) regressor 
the depth of which is {$10$} and the width of which is $24$ with the adaptive moments variant of stochastic gradient descent to optimize the mean squared error loss function.  {We use rectified linear activation functions in the hidden layers.}  We train the model using the \textit{entire} data to capture all variance in the data and then plot.

\vspace*{-1em}
\subsection{Performance Measures}
Measuring the performance of the two proposed approaches is challenging due to their unsupervised nature.  However, using the expert's input $i_\text{true}^\ast$ and the output $i^\ast$ from \eqref{eq:rca}, we define an accuracy measure for RCA as:
\begin{equation}
    \mathsf{Accuracy} \coloneqq \frac{1}{m} \sum \mathbbm{1}[i^\ast = i^\ast_\text{true}].
    \label{eq:accuracy}
\end{equation}%

For relationship discovery, a similar measure does not exist; however, a check that plots do not violate known upper bounds or subject matter expertise understanding can be used, especially in the presence of the DNN plotted regression.%

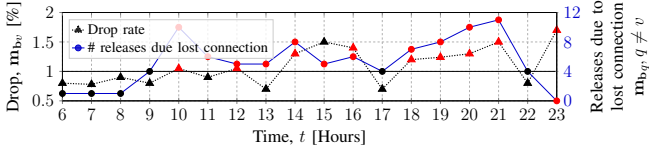
\begin{figure}[!t]
\centering
\resizebox{0.48\textwidth}{!}{\begin{tikzpicture}
\tikzstyle{every node}=[font=\large,scale=2]


\begin{axis}[
width=10in,
height=2.3in,
tick align=outside,
tick pos=left,
every y tick label/.append style={black},
x grid style={white!69.01960784313725!black, dashed},
xlabel={Time, $t$ [Hours]},
xmin=6, xmax=23,
xtick={6,7,...,23},
y grid style={white!69.01960784313725!black, dashed},
ylabel={Drop, $\mathbf{m_b}_v$ [\%]},
ymin=0.5, ymax=2.0,
ytick={0.5,1.0,...,2.5},
grid=both,
scatter/classes={%
    a={black},%
	b={red}}
]

\addplot [line width=1.6pt, dotted]
table {%
6 0.8
7 0.78
8 0.9
9 0.8
10 1.05
11 0.9
12 1.05
13 0.7
14 1.3
15 1.5
16 1.4
17 0.7
18 1.2
19 1.24
20 1.3
21 1.5
22 0.8
23 1.7
};

\addplot [thick, black]
table {%
5.15 1
23.85 1
};

\addplot [only marks,
  scatter,
  scatter src=explicit symbolic,
  mark=triangle*, mark size=6, mark repeat=1, mark options={solid}
]
table [meta=label]{%
x  y  label
6 0.8 a
7 0.78 a
8 0.9 a
9 0.8 a
10 1.05 b
11 0.9 a
12 1.05 b
13 0.7 a
14 1.3 b
15 1.5 a
16 1.4 b
17 0.7 a
18 1.2 b
19 1.24 b
20 1.3 b
21 1.5 b
22 0.8 a
23 1.7 b
};
\end{axis}

\begin{axis}[
width=10in,
height=2.3in,
axis y line=right,
axis line style={-},
every axis label/.append style ={black},
every tick label/.append style={black!20!blue},
legend cell align={left},
legend entries={\smaller Drop rate,{\smaller \# releases due lost connection}},
legend style={
  fill opacity=0.8,
  draw opacity=1,
  text opacity=1,
  at={(0.22,0.92)},
  anchor=north,
  draw=white!80!black,
  font=\tiny
},
xtick={6,7,...,23},
xticklabels={,,},
xmin=6, xmax=23,
ylabel style={align=center},
ylabel={Releases due to \\ lost connection \\ $\mathbf{m_b}_q, q\neq v$},
ymin=0, ymax=12,
ytick={0,4,...,12},
ytick pos=right,
scatter/classes={%
    a={black},%
	b={red}}
]
\addlegendimage{black,line width=1.6, dotted,mark=triangle*,mark size=6}
\addlegendimage{black!20!blue,mark=*,mark size=4}
\addplot [line width=1.2pt, black!20!blue] 
table {%
6 1
7 1
8 1
9 4
10 10
11 6
12 5
13 5
14 8
15 5
16 6
17 4
18 7
19 8
20 10
21 11
22 4
23 0
};

\addplot [only marks,
  scatter,
  scatter src=explicit symbolic,
  mark=*, mark size=4, mark repeat=1, mark options={solid}
]
table [meta=label]{%
x  y  label
6 1 a
7 1 a
8 1 a
9 4 a
10 10 b
11 6 b
12 5 b
13 5 b
14 8 b
15 5 b
16 6 b
17 4 a
18 7 b
19 8 b
20 10 b
21 11 b
22 4 a
23 0 b
};

\end{axis}
\end{tikzpicture}}%
\vspace*{-1em}\caption{Detecting anomaly (red {dots}) in a KPI {(black)} and one of its building counters {(blue)} over a period of time where $\vert\phi\vert \approx 0.7071$ and a $1$\% threshold.}
\label{fig:an_anomaly}
\end{figure}%

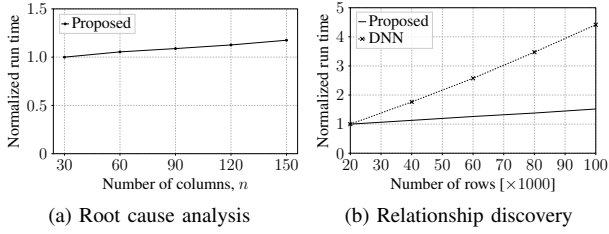
\begin{figure}[!t]
\centering
\subfloat[Root cause analysis]{
\resizebox{0.22\textwidth}{!}{
\begin{tikzpicture}
\tikzstyle{every node}=[font=\Large,scale=2.5]
\begin{axis}[
width=8in,
height=5in,
legend cell align={left},
legend style={
  fill opacity=0.8,
  draw opacity=1,
  text opacity=1,
  at={(0.18,0.83)},
  anchor=south,
  draw=white!80!black,
  font=\tiny
},
tick align=outside,
tick pos=left,
x grid style={white!69.0196078431373!black, dashed},
xlabel={Number of columns, $n$},
xmajorgrids,
xmin=24, xmax=156,
xtick style={color=black},
xtick={30,60,...,150},
xticklabels={
  \(\displaystyle {30}\),
  \(\displaystyle {60}\),
  \(\displaystyle {90}\),
  \(\displaystyle {120}\),
  \(\displaystyle {150}\),
},
y grid style={white!69.0196078431373!black, dashed},
ylabel={Normalized run time},
ymajorgrids,
ymin=0, ymax=1.5,
ytick style={color=black},
ytick={0,0.5,...,1.5},
yticklabels={
  \(\displaystyle {0}\),
  \(\displaystyle {0.5}\),
  \(\displaystyle {1.0}\),
  \(\displaystyle {1.5}\)
},
]
\addplot [ultra thick, black, mark=*, line width=1.2pt]
table {%

30 1
60 1.05457447987934
90 1.08878286867723
120 1.12636417528794
150 1.1751349029768
};
\addlegendentry{Proposed}
\end{axis}

\end{tikzpicture}}\label{fig:runtimes_rca}}
\subfloat[Relationship discovery]{ 
\resizebox{0.22\textwidth}{!}{
\begin{tikzpicture}
\tikzstyle{every node}=[font=\Large,scale=2.5]
\begin{axis}[
width=8in,
height=5in,
legend cell align={left},
legend style={
  fill opacity=0.8,
  draw opacity=1,
  text opacity=1,
  at={(0.18,0.72)},
  anchor=south,
  draw=white!80!black,
  font=\tiny
},
tick align=outside,
tick pos=left,
x grid style={white!69.0196078431373!black, dashed},
xlabel={Number of rows [$\times 1000$]},
xmajorgrids,
xmin=20, xmax=100,
xtick style={color=black},
xtick={10,20,30,40,50,60,70,80,90,100,110},
xticklabels={
  \(\displaystyle {10}\),
  \(\displaystyle {20}\),
  \(\displaystyle {30}\),
  \(\displaystyle {40}\),
  \(\displaystyle {50}\),
  \(\displaystyle {60}\),
  \(\displaystyle {70}\),
  \(\displaystyle {80}\),
  \(\displaystyle {90}\),
  \(\displaystyle {100}\),
  \(\displaystyle {110}\)
},
y grid style={white!69.0196078431373!black, dashed},
ylabel={Normalized run time},
ymajorgrids,
ymin=0, ymax=5,
ytick style={color=black},
ytick={0,1,2,...,5},
yticklabels={
  \(\displaystyle {0}\),
  \(\displaystyle {1}\),
  \(\displaystyle {2}\),
  \(\displaystyle {3}\),
  \(\displaystyle {4}\),
  \(\displaystyle {5}\)
},
]
\addplot [ultra thick, black]
table {%
20 1
40 1.13
60 1.262281685669
80 1.38137498666236
100 1.52052973186797
};
\addlegendentry{Proposed}
\addplot [ultra thick, black, dashed, mark=x, mark size=6, mark options={solid}]
table {%
20 1
40 1.76365
60 2.57516
80 3.47185
100 4.414863
};
\addlegendentry{DNN}
\end{axis}

\end{tikzpicture}}\label{fig:runtimes_reldisc}}
\caption{Run time: a) for RCA as a function of columns $n$ and b) for relationship discovery and DNN regression as a function of rows $m$.}
\label{fig:runtimes}
\end{figure}%

\begin{figure}[!t]
\centering
\subfloat[Uplink SE vs SINR]{\resizebox{0.22\textwidth}{!}{\includegraphics{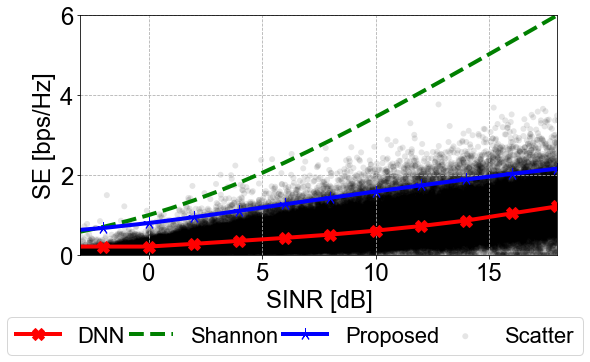} \label{fig:closed_form}}}
\subfloat[Traffic volume vs users]{\resizebox{0.22\textwidth}{!}{\includegraphics{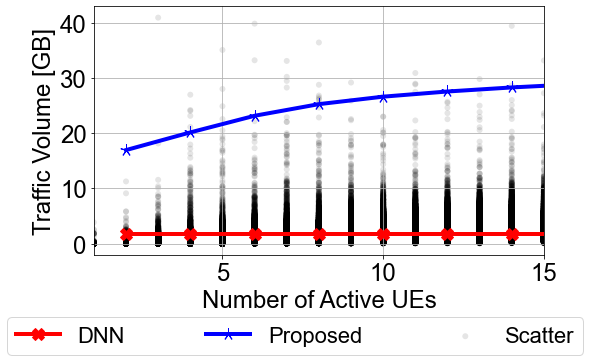}\label{fig:max_agg}}}
\caption{Relationship discovery vs deep neural network regression: (a) known closed-form upper bound solution and (b) unknown closed-form solution.}
\label{fig:rel_discoveries}
\end{figure}%

\vspace*{-1em}\subsection{Discussion}
We start the discussion with the RCA algorithm. The results in Table~\ref{table:performance} shows the accuracy of RCA as computed using \eqref{eq:accuracy}.  The accuracy is low in Scenario A since {the number of data points $5$ is too small to construct density regions and correctly detect anomalies. However, this accuracy increases in Scenario B as the number of data points increases making the formation of such regions more likely}. {We show detected anomalies in our studied KPI at $1$\% and one counter in Fig.~\ref{fig:an_anomaly}}.  

Fig.~\ref{fig:runtimes_rca} shows that RCA runs in $\mathcal{O}(n)$.  However, with a fixed size $\mathbf{M}$ per BS for every time window, it becomes $\mathcal{O}(1)$.

Next, for the relationship discovery algorithm in Fig.~\ref{fig:rel_discoveries}, we observe that the known theoretical upper bound relating the SINR with the SE \eqref{eq:shannon} is not being violated by the plot generated by the algorithm in Fig.~\ref{fig:closed_form}.  Further, when the relationship involves a discrete quantity, such as the number of users shown in  Fig.~\ref{fig:max_agg}, we observe the poor performance of the DNN regressor as it underfits the data making our relationship discovery application suitable for such cases.

Relationship discovery runs in near-constant time as shown in Fig.~\ref{fig:runtimes_reldisc}.  Contrast this with the DNN the run-time complexity of which is linear in the number of data points (with much steeper slope). The reason for this slight deviation from constant run-time is due to the strict convergence criterion, which intuitively depends on the number of data points.

An insight from this discussion is that this near-constant run-time complexity combined with subject-matter expert validation checkpoints can make unsupervised learning suitable for real-time diagnosis.  These checkpoints serve as an opportunity to re-adjust the hyperparameters of models and also a sanity check against unexpected outcomes.

\section{Conclusion}\label{sec:conclusion}

In this letter, we proposed two simplified applications of unsupervised learning for the use of real-time performance self-diagnosis in next-generation RAN.  The algorithms behind these two applications run in near-constant time making them suitable for real-time analysis.  As in any unsupervised learning algorithm, the subject-matter expert input is needed to validate outputs.  An interesting extension is to connect these applications with means to enable networks to perform autonomous optimization of 
service-related objectives (e.g., throughput or latency) or radio resource management related objectives (e.g., handovers and resource allocation). This is an idea that can not only enable intelligent self-diagnosis in real-time RIC in 5G, but even in 6G and beyond.

\bibliographystyle{IEEEtran}
\bibliography{main.bib}

\end{document}